# A Human Digital Twin Architecture for Knowledge-based Interactions and Context-Aware Conversations

Abdul Mannan Mohammed, Azhar Ali Mohammad, Jason A. Ortiz, Carsten Neumann, Grace Bochenek,
Dirk Reiners, Carolina Cruz-Neira
University of Central Florida
Orlando, Florida
abdulmannan.mohammed@ucf.edu, azharali.mohammad@ucf.edu, jason.ortiz@ucf.edu,
carsten.neumann@ucf.edu, grace.bochenek@ucf.edu, dirk@ucf.edu, carolina@ucf.edu

**ABSTRACT**

Recent developments in Artificial Intelligence (AI) and Machine Learning (ML) are opening novel opportunities for human-autonomy teaming (HAT) to accomplish a task, a mission, or a continuous coordinated activity. The challenge is to provide awareness and control to the humans over autonomous assets and their actions, while having trusted interactions with them as teammates and supporting HAT shared contextual understanding to accomplish the task. Addressing this challenge is crucial for the success of hybrid human-autonomous teams pursuing a common goal. To address this challenge, we present a real-time Human Digital Twin (HDT) software architecture that integrates Large Language Models (LLM) focused on knowledge reporting, answering, and recommendations, into a visual interface that provides a life-like physical embodiment of the autonomous system. We use a metacognition approach to empower the LLM to have a deeper and personalized understanding of the human (or humans) that it needs to interact with, so it can provide context-aware responses aligned with the human's expectations and needs. Our HDT(s) then becomes a visually and behaviorally recognized team member(s) that can be integrated into the complete life cycle of a mission, from training to deployment to after-action-review.

We present an open architecture that provides a protocol to integrate customized LLMs to enhance dialogue quality and increase context sensitivity. This architecture encompasses sophisticated speech recognition, context-aware processing for adaptive learning responses, AI-driven dialogue generation, AI emotion engine, lip-syncing, and lifelike visual and auditory feedback. With this architecture, the HDT can do real-time interactions without needing explicit dialogue in advance, capturing multimodal data to create a realistic in-context conversation.

This paper describes the HDT system architecture and its performance metrics, highlighting critical functionalities developed and opportunities for further development. Our HDT is targeted to support HAT through personalized interactions, heightened realism, and adaptability to context of operation.

**ABOUT THE AUTHORS**

**Abdul Mannan Mohammed** is a second-year master's student in Computer Engineering at the University of Central Florida (UCF). He received his B.S. in Computer Science and Engineering from Osmania University, India. Currently, he works at Virtual and Augmented Reality Applications Lab (VARLab) as a graduate research assistant. He has previously served as the Video Chair for IEEE VR (Virtual Reality) 2024. His research interests lie in the fields of human-autonomy teaming, embodied agents, virtual humans, user experience, and usability.

**Azhar Ali Mohammad** is a second-year student at UCF pursuing a master's degree in computer engineering. He graduated from Osmania University in India with a bachelor's degree in Computer Science and Engineering. Currently, he works as a graduate assistant at the VARLab. Azhar is mainly focused on innovative projects that merge artificial intelligence with virtual reality. He is profoundly involved in human-computer interactions, specifically concentrating on the development of human digital twins.

**Jason A. Ortiz** is a Modeling & Simulation Ph.D. student at UCF conducting research on remote collaboration in extended reality environments at the VARLab.

**Carsten Neumann** is a Research Associate in Computer Science at UCF and a research scientist with the VARLab group led by Dr. Carolina Cruz-Neira. He has over 20 years of experience in immersive environments, software architectures design, and real-time rendering and computing. Previously, Carsten was the Senior Research Scientist at the Emerging Analytics Center (EAC) which he joined in 2014. Before joining the EAC, Carsten was a research scientist at the Louisiana Immersive Technologies Enterprise and the University of Louisiana at Lafayette.

**Grace Marie Bochenek** is director of the School of Modeling, Simulation and Training, and the Institute for Simulation and Training at the University of Central Florida. She is former Director of the National Energy Technology Laboratory within the United States Department of Energy's Office of Fossil Energy. She also served as the acting United States Secretary of Energy in early 2017. She previously had spent much of her career at the U.S. Army Tank Automotive Research, Development and Engineering Center.

**Dirk Reiners** is an Associate Professor in the department for Computer Science at the University of Central Florida. His main research interests are in VR/AR and immersive displays and software systems to drive them, and applications thereof including immersive data visualization and Digital Twins. Before joining academia, he worked for more than 10 years at Fraunhofer IGD, a private research institute and the largest research group for Computer Graphics in the world. He was the lead designer and project lead of the OpenSG Open Source scenegraph project and received an IITSEC Best Paper award. He is a member of IEEE and ACM.

**Carolina Cruz-Neira** (National Academy of Engineering member) is the Agere Chair at the Computer Science Department at the University of Central Florida, Orlando, FL, USA. She is a XR and Modeling & Simulation Hall of Fame pioneer in the areas of virtual reality and interactive visualization, having created and deployed a variety of technologies that have become standard tools in industry, government, and academia. She is known worldwide for being the creator of the CAVE virtual reality system. Dr. Cruz-Neira is an IEEE Fellow, member of the IEEE Virtual Reality Academy and has received the IEEE Virtual Reality Technical Achievement Award, among many other recognitions.

# A Human Digital Twin Architecture for Knowledge-based Interactions and Context-Aware Conversations

## INTRODUCTION

The collaboration between humans and intelligent, autonomous agents, known as human–autonomy teams (HATs), presents significant challenges and promising opportunities. Establishing trust, maintaining control, and ensuring shared situational awareness are critical for HATs, especially in specialized settings like healthcare, military operations, and disaster response (Wynne & Lyons, 2018). The rapid expansion of artificial intelligence (AI) technologies has introduced new interaction realms necessitating sophisticated coordination between human and AI team members (Jiang et al., 2022). This dynamic interdependence requires continuous learning and adaptation to enhance team performance (O'Neill et al., 2022). Mutual trust can be fostered through transparency, explicit explanations, and consistent AI performance (Murdick, 2023).

Advancements in AI have unlocked new possibilities for humans to collaborate more effectively with autonomous assets. AI technologies now allow for the deployment of intelligent systems capable of performing cognitive tasks traditionally handled by humans, such as learning, interacting, solving problems, and making decisions (Dellermann et al., 2021). These advancements have enabled significant leaps forward in simulating human teammates (e.g., Manfredi et al., 2023). HATs, with their efficiency, precision, and situational awareness, have great potential across various fields. AI-powered context-aware services heavily rely on human-AI cooperation; while AI systems can obtain and analyze contextual data, human interaction is necessary to verify and improve these insights to ensure their relevance and correctness (Jiang et al., 2022). We anticipate that, in certain contexts, the autonomous assets need to provide a sense of physical presence with a look and feel resembling a real human for better acceptance, trust, and interactions. This need has driven the design and development of a real-time Human Digital Twin (HDT) software architecture that integrates Large Language Models (LLM) focused on knowledge reporting, answering, and recommendations, into a visual interface that provides a life-like physical embodiment of the autonomous system.

Until recently, the term HDT was applied to a digital recreation of a human, which enables real-time monitoring of and synchronization with physical counterparts (Shengli, 2021). In engineering, HDT have also been used to address human-centered design and manufacturing, with the goal of a paradigm shift in human-system integration by coupling humans' characteristics directly to the system design and its performance. These applications for HDT are proven beneficial and are rapidly advancing a broad range of disciplines, however, humans are complex physical, cognitive, and emotional beings that interact with each other and their surrounding environment; they have roles in the different situations and contexts they encounter; they communicate with other humans around them and develop many types of relationships. Research to address these aspects and how to reflect those in HDTs is still in its early stages and it is the focus of this paper. We believe that we need to incorporate cognitive and mental capabilities in our system model. To address this, the HDTs need to be able to acquire information, process it, and use it for problem solving. Metacognitive AI systems can help in achieving this goal, as they can support the HDTs to better predict and respond to human requirements by understanding mental states and cognitive resources (Ackerman & Thompson, 2017). The integration of multimodal data, such as visual, auditory, and contextual input, further enhances the capability of AI systems to provide contextually appropriate and emotionally resonant responses. However, this integration presents challenges, including the need for sophisticated data processing and the management of potential inconsistencies in AI behavior as well as an optimized system of systems software framework to enable the real-time performance and rendering of the HDT.

In summary, this paper presents an open software architecture for HDTs that incorporates Large Language Models (LLMs) to enhance the interaction between humans and AI through enhanced dialogue quality and increased context sensitivity. By employing a metacognitive strategy and integrating multimodal data, the HDT provides personalized interactions, heightened realism, and adaptability to context of operation.

## RELATED WORK

Our work presents a new approach to HDT by incorporating LLM, AI, metacognition, and real-time photo realistic image generation to enable in-context, voice-based, natural, novel conversations to create a HDT with a credible personality, memory, behavior, and appearance. Our system uniquely integrates multiple advanced technologies and design principles into an open architecture to enhance HATs collaboration. Therefore, in this section, we focus on related works that pertain to each component of our system, highlighting how they contribute to the overall architecture and objectives of our HDT implementation. This approach allows us to contextualize our contributions within the broader landscape of existing technologies and methodologies.

**Conversational Agents**
Conversational agents have evolved from basic rule-based systems (Hussain & Sianaki, 2019) to advanced AI models, such as OpenAI's GPT-3 and GPT-4, which are LLMs. These models greatly improve the capacity to produce coherent and context-aware responses. There has been an emphasis on the potential of Virtual Humans (VHs) that incorporate LLMs to evoke social emotions and facilitate realistic interactions (Llanes-Jurado et al., 2024). These VHs integrate realistic avatars with sophisticated communication systems that incorporate psychological elements like personality, mood, and attitudes (So et al., 2023). This integration improves the authenticity and realism of interactions. To enable more significant and emotionally impactful interactions with conversational agents, we can turn to affective computing, a field that centers on the identification and reaction to human emotions (Arya et al., 2021). Affective computing techniques allow to train autonomous systems with more human-like abilities by providing them with physiological and behavioral information such as facial images, body posture, voice, as well as heart rate, skin temperature to name a few. Conversational agents incorporating affective computing seem to provide the desired affective skills, although the need for these depends largely on the application's context (Hernandez et al., 2023).

**Generative Pre-training Transformer (GPT) over Manual Information Search Tasks**
Research suggests that GPT models outperform manual information searches in terms of their effectiveness and ability to rapidly process substantial amounts of data. GPT models, such as ChatGPT, produce responses that are contextually appropriate and logically connected, hence reducing the need for human intervention and expediting the process. GPT-4 possesses excellent qualities that enable it to rapidly amalgamate information from multiple sources, generating comprehensive and precise outcomes that manual information searching might fail to identify (Sandmann, 2024). ChatGPT has been shown to summarize and answer complicated queries, and automate repetitive data entry operations, improving efficiency and accuracy (Ray, 2023). Integrating generative AI into workflows is revolutionary, improving output speed and quality beyond manual searching and data processing (Alawida et al., 2023).

**Chat GPT-4o**
GPT-4Vision, developed by OpenAI, is a breakthrough in incorporating multimodal functionalities into LLMs, allowing them to interpret and react to visual inputs in real-time (OpenAI, 2023; Gallagher & Skalski, 2023; Guzman, 2024). This enables applications like optical character recognition (OCR), object detection, and image captioning, hence improving the practical usefulness of ChatGPT. In May of 2024, GPT-4o was released, including enhanced contextual understanding and improved efficiency and speed (OpenAI, 2024). OpenAI shows the integration of advanced natural language processing capabilities ensures that GPT-4o can handle complex queries with greater ease and precision.

**Metacognitive AI**
The understanding and management of one's own thoughts and behaviors, known as metacognition, is essential in the development of efficient Generative AI (GenAI) systems. LLMs, such as GPT-4, pose substantial usability issues in GenAI, related to increased mental loads for users about how they find, use and validate information coming from the models. To effectively use these systems, users need high levels of metacognitive monitoring and control. This is because users need to comprehend and assess outcomes, prompt efficiently, and integrate these technologies into their workflows. Recent research publications suggest that incorporating metacognitive support tactics into GenAI systems can reduce these requirements by improving the ability to provide explanations and customization options (Tankelevitch et al., 2024). Therefore, as we look into the future of HDTs, metacognition-driven approaches become necessary, in particular to address issues related to reliability and trust. Our HDT uses metacognitive AI by integrating visual and auditory inputs to enhance context-aware, emotionally sensitive, and realistic user interactions.

**Personality, Memory, and Behavior of LLMs in Human Digital Twin Systems**
The personality and behavior of LLMs play a vital role in HDT systems since they have a substantial influence on the effectiveness and credibility of these systems in imitating human-like interactions. Studies have demonstrated that LLMs, such as GPT-3.5 and GPT-4, may be programmed to display personality characteristics like those in the Big Five personality model (Roccas et al., 2002). The speech patterns and behaviors demonstrated by the LLMs are influenced by these attributes, resulting in more personalized and relatable interactions with users. LLMs designed to cater to extroverted individuals may employ more emotive and captivating language, hence boosting the user's overall experience (Jiang et al., 2024). However, there are limitations, most notably the inconsistency in the quality of LLM output, since LLMs frequently struggle to continuously follow instructions or maintain proper personality attributes during interactions. This ultimately results in decreased trustworthiness and coherence. By saving previous conversation history and recalling personality attributes throughout interactions, our HDT system maintains personality traits effectively across sessions.

**AN ARCHITECTURE FOR HUMAN DIGITAL TWIN**
Our HDT technology makes use of context-aware processing, an AI emotion engine, and sophisticated AI-driven dialogue generation to guarantee real-time interactions that preserve emotional coherence and facilitate seamless dialogues. Our HDT system utilizes multimodal data capture and adaptive learning responses to facilitate seamless, lifelike, and emotionally resonant encounters. Furthermore, our architecture incorporates a real-time visual interface that provides a life-like physical embodiment of the HDT in context with the conversation. This effectively addresses the challenges encountered with standard conversational agents.

Engaging the user in a convincing and conversational manner is achieved by utilizing advanced language models such as ChatGPT-4. The HDT system architecture comprises these essential components: A) a Visual interface to capture the human's conversation and deliver the HDT responses through the Speech-to-Text Module, Animation Module, Text-to-Speech Module, Lip Sync Generation; and B) an AI module that captures non-verbal user interactions and processes the appropriate responses through a Visual Input Capture, and a LLM model. Every individual component has a distinct function in the system. Figure 1 illustrates the overall interaction flow and design of the HDT system.

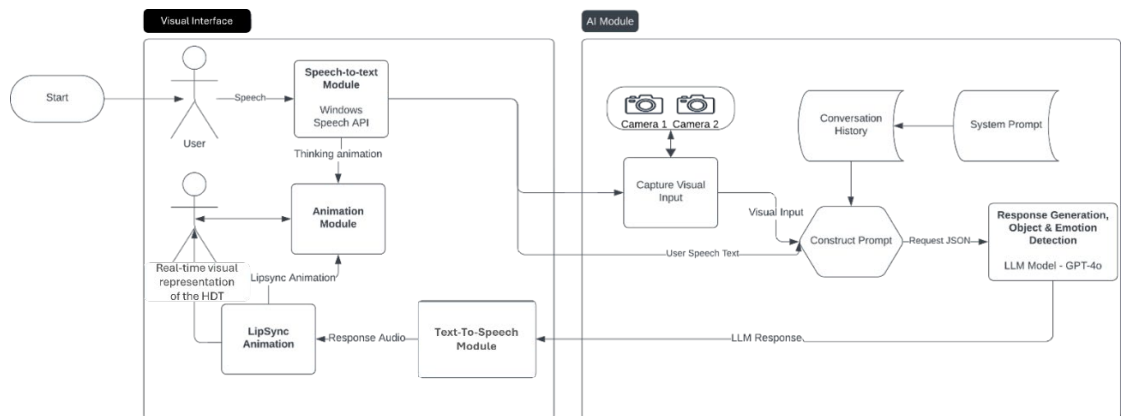

**Figure 1. HDT System Design**

**A. Visual Interface**
The HDT is designed to operate in a variety of presentation modes to the user, that is, the HDT can be presented on a conventional monitor, can be embedded in a virtual reality interface, a holographic display or other visual platforms required to interact with it. The core is a real-time rendering engine capable of creating the digital human form for the HDT integrating all the elements needed for the conversation "believability": Looks, facial expressions, lip movements, gaze, eye blink, clothing, background environment and others. For our prototype design we use Unreal Engine 5.2.1 as the central setting in which human/HDT interactions occur. For the HDT visual representation we use the Unreal Metahuman personas with an interface that includes a microphone and a set of speakers to capture the user's conversation and deliver the HDT responses generated by the AI module. We define a set of interoperability protocols among all the modules for seamless integration of their functionality and therefore provide immediate visual and auditory responses to the user through the HDT.

**Speech-to-Text Module:** The Speech-to-Text Module converts the user's speech into written text. The functionality of this module is crucial as it converts the auditory information provided by the user into a format that can be effectively analyzed and interpreted by the AI module. As the user speaks, their speech is recorded and transcribed, which then initiates a thinking animation in the HDT to visually indicate that the system is processing the information. For our prototype implementation of the architecture, we chose the Windows Speech API (Microsoft, n.d.) as it being used extensively across a broad range of users.

**Animation Module:** The Animation Module is tasked with triggering suitable body and face movements for the HDT as well as seamless transitioning from previous movements and expressions to provide the illusion of a living human. It also incorporates reflex movements, such as eye blinking, breathing, and "fiddling" such as small head or hand movements. The system integrates information from both the Speech-to-Text and the Text-to-Speech Modules to coordinate the character's movements with the speech input and output as well as feedback from the metacognition inputs. This also involves triggering lip-sync animations to ensure enhanced realism and consistency in mouth movements during the interaction.

**Text-to-Speech Module:** The text-to-speech transforms textual responses from the AI module into an appropriate audible voice for the HDT. This module produces the auditory output to enable the HDT participating in a fluid conversation with the user. For our prototype, we use the Unreal Speech service (*Text-to-Speech API*, n.d.).

**Lip Sync Animation:** An element that adds significant acceptance of the HDT is its ability to move its lips synchronized with the text-to-speech output. We incorporated a broad range of lip animations integrated into the face of our HDT to provide realistic lip synchronization based on audio inputs generated by the AI module. Our prototype uses the Unreal Engine MetaHumanSDK.

**Visual Input Capture Module:** The HDT is equipped with cameras to capture visual input from the user and the user's environment. This visual data is processed and sent to the AI module, where it is used to enhance the contextual understanding of the interaction. This is the approach to incorporate metacognition into our work, as we can analyze these visual cues, to interpret the user's intentions and emotions, as well as objects and environments around the user, which lead to more accurate and contextually appropriate responses.

**B. AI Module**

The AI Module is the core of the HDT system, which for our prototype is built with ChatGPT-4o (OpenAI, 2024). This module handles several critical tasks: (a) **Conversation History**: Maintains a log of previous interactions to ensure continuity and context in the ongoing conversation; (b) **System Prompt**: Sets the initial context and parameters for the AI's responses; (c) **Construct Prompt**: Combines user speech, text and visual input to create a comprehensive prompt for the AI; (d) **Response Generation, Object & Emotion Detection**: Utilizes ChatGPT-4o's capabilities to generate responses that are contextually appropriate and emotionally resonant. The AI module also detects objects and emotions from the visual input to refine its responses further.

**Interaction Flow**

The user/HDT interaction flow is defined as follows: (1) **User Interaction**: The user starts by speaking to the HDT system, which is captured by the Speech-to-Text Module and the Visual Input Capture Module; (2) **Processing Input**: The text and video stream is sent to the AI module; (3) **AI Response**: The AI module processes the input, referencing the conversation history and system prompt to generate an appropriate response; (4) **Output Generation**: The response text is sent to the Text-to-Speech Module to generate the audio output, which is then synchronized with the lip-sync animations; (5) **HDT Generation**: The Animation Module ensures that the HDT visual persona displays the appropriate motions, facial expressions, and lip synching during the interaction, as well as the continuous real-time flow delivered to the user to support the realism of the experience. By integrating these components, the HDT system offers a comprehensive and immersive user interaction experience to users. The synergy between the visual, auditory, and contextual processing capabilities of the HDT ensures that the user can engage with their digital twin in a manner that feels natural and intuitive.

**HDT SYSTEM PERFORMANCE ANALYSIS SET UP**

In this section we present a preliminary assessment of the HDT system's performance in supporting human-autonomy teaming (HAT) within a controlled environment. We chose a gun assembly task as it has well-defined steps and constraints. As system designers, we defined a set of sequential sub-tasks for the gun assembly and in each sub-tasks a set of most-likely user questions, comments, and request for more information. Our goal was to assess the HDT's ability to detect and communicate anomalies during assembly, such as missing or incorrectly assembled parts, not to assess how human users interact with the HDT. This assessment helps determine the system's capability to assist in complex tasks by providing accurate, context-aware feedback and proactive problem-solving insights, ultimately enhancing task efficiency, reducing errors, and improving user experience.

**HDT Physical Embodiment Setup**

For this assessment, we chose a no-glasses autostereoscopic 55" display (Magnetic3D DS) instrumented with a microphone, speakers, and two webcams to present the physical appearance of the HDT to the user and to capture user's input and actions. Gun parts are laid out on a table in front of the display within the camera's field of view, arranged to replicate a typical assembly setup (Figure 2). One of the two high-definition web cameras capture detailed visual input of the gun parts, and the other focuses on the user to detect emotion and capture non-verbal cues. Prior to starting the conversation, the HDT system receives comprehensive information about the gun assembly process, including a digital manual with detailed diagrams and step-by-step instructions (Figure 3). The system is configured to respond to specific user queries and recognize various gun components and tools. It processes visual and auditory inputs simultaneously, allowing it to provide context-aware feedback.

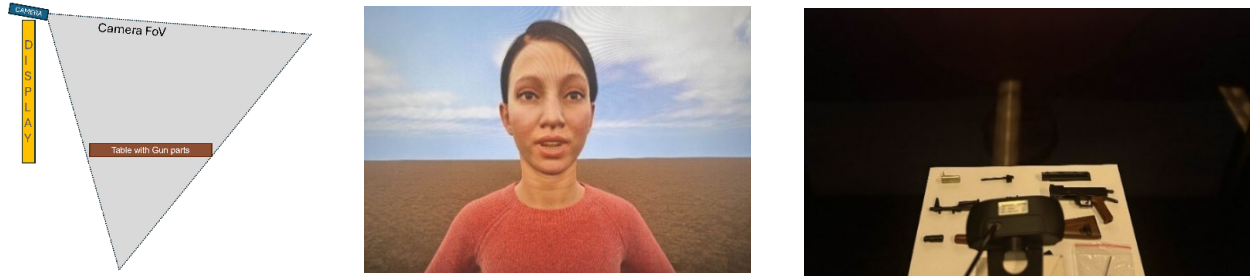

**Figure 2. HDT System: System components (left) HDT on Display (middle) and Gun Assembly Station (right)**

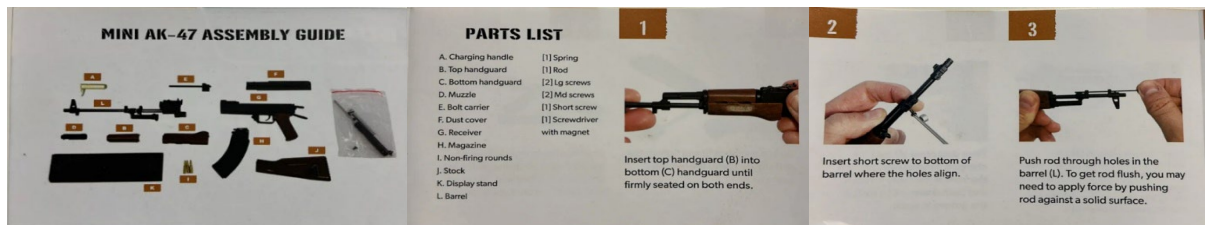

**Figure 3. Gun Assembly Digital Manual**

**Assessment Procedure**

We divided the gun assembly task into the five phases described below. Within these categories, we defined a minimum of six different scenarios that the HDT may encounter when interacting with human users. Examples of these scenarios are: Not all gun parts are present when assembly task is initiated, HDT should provide feedback on what parts may be missing; the user may ask the HDT for verification of having selected the correct part; clarification on how to execute one of the assembly steps; recommendations for next course of action after a mistake; emotional encouragement if the user is not confident about his/her performance; feedback on correct completion of the task. The goal of these scenarios is to provide insights into the system's real-time visual analysis, context-aware feedback, metacognition, and problem-solving capabilities. This structured approach allows us to comprehensively evaluate the HDT system's performance across various dimensions of human-autonomy teaming, to verify that both humans and AI can work together effectively to complete the task and achieve the goal.

**Phase 1: Pre-assembly and Initial Checks:** In this phase, the focus is on real-time visual analysis and context-aware feedback to the user from the HDT. The HDT system is tasked with verifying all parts' presence and identifying specific components upon request. This involves using the visual capabilities of GPT-4o for HAT to scan and recognize each part laid out for assembly. The system provides real-time feedback by visually identifying and confirming parts such as the receiver, charging handle, and muzzle. This ensures that all necessary components are present and correctly identified before the assembly process begins, reducing the risk of errors caused by missing or misidentified parts (Figure 4).

**Phase 2: Instructional Guidance:** During the instructional guidance phase, the HDT system employs context-aware feedback and Problem Solving to assist users through the assembly steps. The system breaks down complex instructions into manageable sub-steps and provides detailed explanations when users are confused about specific procedures. For example, it should elaborate and guide users on how to properly load and click the magazine into place. This phase highlights the system's ability to offer precise, step-by-step guidance, ensuring that users understand and correctly execute each part of the assembly process, thereby enhancing the efficiency and accuracy of the task.

**Phase 3: Recommendations/Solutions:** The HDT system leverages its problem solving and metacognition capabilities to offer recommendations and solutions for issues encountered during assembly. When users make mistakes, such as using the wrong screw or facing obstructions, the system provides corrective actions and alternative approaches. For example, it suggests a gentler method for addressing threading issues or advises on the safe continuation despite a seemingly incorrect screw length. This proactive problem-solving approach helps users navigate unexpected challenges, ensuring that the assembly process can proceed smoothly even when deviations occur.

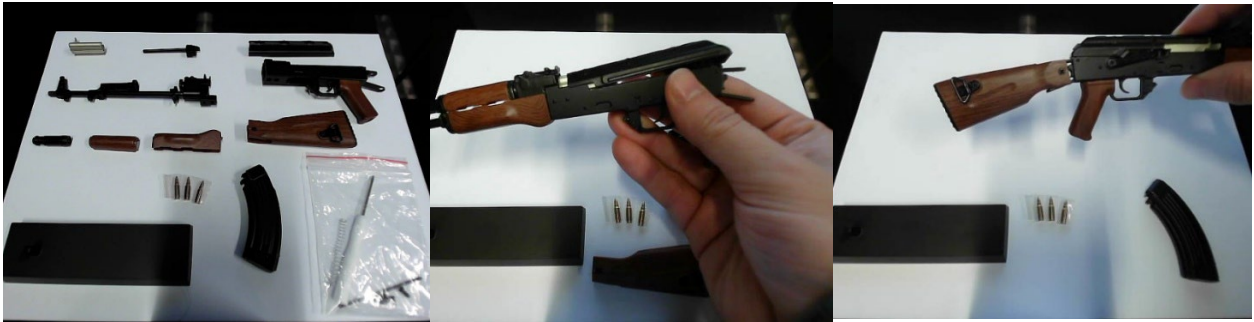

**Figure 4. Gun Assembly Phases During Assembling Process**

**Phase 4: Emotional Support:** The emotional support phase focuses on context-aware feedback and metacognition to address the user's emotional and motivational needs. The HDT system detects emotional cues through visual and auditory inputs, including facial expressions and spoken words, and responds with appropriate support. It offers encouragement, reassurance, and tips for calming down when users feel frustrated or anxious. For instance, if a user feels overwhelmed at any of the assembly steps, the system provides words of encouragement and suggests techniques to improve focus and reduce anxiety. By addressing the emotional aspects of the assembly task, the HDT system helps maintain user motivation and confidence, contributing to a more positive and productive experience.

**Phase 5: Assembly Verification:** Here, the HDT system utilizes real-time visual analysis and context-aware feedback to ensure that the assembly has been completed correctly. The system checks the installation and alignment of components such as screws, the muzzle, and the dust cover, providing immediate feedback on any discrepancies. For example, it verifies that the spring is properly positioned on the bolt carrier and that the stock is correctly oriented. This phase ensures that the final assembly meets the required standards, identifying any missed steps or errors before the task is considered complete. By doing so, the system enhances the reliability and safety of the assembled product. The goal of these phases is to facilitate HAT, where humans depend on AI for guidance and motivation, and AI relies on human input to complete tasks, working together to achieve the overall goal.

**RESULTS**

This section discusses the resulting performance of the HDT during a gun assembly task following the set up described in the previous section. We collected the HDT responses and accuracy of such responses to determine the HDT's success/failure at each phase. The results are summarized in Table 1, highlighting the system's success, conditional success, and failure rates. For the HDT performance characterization, we have defined the following: *Success*: The HDT gives the correct response without any user feedback. *Conditional Success*: The HDT initially gives incorrect responses but reaches the correct response after receiving user feedback (up to a maximum of five user feedback attempts). *Failure*: The HDT cannot provide the correct response even after five user feedback attempts. For this assessment, we want to note that the user feedback was a set of conversational paragraphs pre-defined by the designers based on most likely conversations that human users may do to request more detailed responses from the HDT. In the rest of the section, when describing scenarios, we provide the essence of the HDT conversation for the purposes of discussion but note that the HDT provides variations and more detailed conversational paragraphs when interacting with the user[1]. For example, Scenario 1.4 (Let's verify the parts against the assembly guide), it is actually delivered by the HDT to the user as: "*Let's verify the parts against the assembly guide: we should have a charging handle, a top handguard, a muzzle, a… (complete list of parts), and a screwdriver. Based on what I see, the charging handle, bolt carrier, rod, and spring appear to be in the same compartment as the bolt carrier. The top handguard, the muzzle is present… (continues listing what is present). Looks like all parts are present*." All tests were executed by the author.

| Phase | Number of Scenarios | Success | Conditional Success | Failure |
|---|---|---|---|---|
| 1.1. Pre-check | 10 | 5 | 5 | 0 |
| 1.2. Identification | 10 | 10 | 0 | 0 |
| 2. Instructional Guidance | 7 | 7 | 0 | 0 |
| 3. Recommendations/Solutions | 10 | 10 | 0 | 0 |
| 4. Emotional support | 11 | 11 | 0 | 0 |
| 5. Assembly verification | 6 | 3 | 3 | 0 |

**Table 1. Summary of HDT System Task Performance**

---
[1] Contact authors if interested in full training and conversational dataset.

**Phase 1: Pre-assembly and Initial Checks**

*Scenario 1.1: Pre-check*
In the pre-check phase, the HDT was tasked with verifying the presence of all parts required for assembly. Out of ten scenarios, the HDT achieved success in five scenarios, conditional success in five scenarios, and no failures. Table 2 presents the number of feedback attempts required per scenario. The HDT successfully identified missing parts after receiving feedback in Scenarios 1.4 (Let's verify the parts against the assembly guide), 1.6 (You have everything you need to proceed with the assembly), 1.7 (If part is missing, we'll proceed without it for now, but remind me to revisit this at the end of assembly.), 1.8 (All the pieces appear to be accounted for), and 1.9 (Please match them with the assembly guide to ensure all parts are ready before we start) demonstrating the system's capability to improve its performance with minimal guidance. The findings indicate that the HDT successfully verified all parts when no parts were removed. This verification included all required pieces, including small ones such as screws.

| Scenario | 1.1 | 1.2 | 1.3 | 1.4 | 1.5 | 1.6 | 1.7 | 1.8 | 1.9 | 1.10 |
|---|---|---|---|---|---|---|---|---|---|---|
| Number of Feedback Attempts | 0 | 0 | 0 | 2 | 0 | 5 | 4 | 3 | 3 | 0 |

**Table 2. Number of Feedback Attempts for Phase 1.1: Pre-check**

However, when specific parts were deliberately removed, such as the display stand, barrel, receiver, and non-firing rounds, the HDT initially provided incorrect responses, stating that these parts were present when they were missing. The HDT required user feedback to correct its responses. Out of five scenarios which were a conditional success, here the number of feedback attempts varied. For example, in scenarios where the display stand or barrel was removed, the HDT corrected itself after an average of three feedback attempts per scenario. An example of a feedback attempt involved the human prompt, "I feel like something is missing. Check again," to which the HDT responded, "Sure, let's take another look... The display stand is not visibly present. Please check the packaging or other included items." This process highlighted that the HDT could correct its initial mistakes and accurately verify the presence or absence of parts after receiving a few feedback prompts, demonstrating its adaptability and learning capability.

*Scenario 1.2: Identification*
In the identification phase, the HDT was responsible for recognizing and confirming specific components upon request. The HDT achieved a 100% success rate, correctly identifying all parts in all 10 scenarios without the need for feedback. The HDT effectively identified and located specific parts upon request. For example, when asked to identify the receiver, the HDT correctly pointed out that it was labeled as part G and showed its position in the parts layout as the component with the wooden grip. Similarly, when asked about the charging handle, the HDT identified it as part A and located it as the small black component with an angled end and rod. This indicates the HDT's strong ability in component recognition, essential for accurate assembly tasks.

**Phase 2: Instructional Guidance**
During the instructional guidance phase, the HDT provided step-by-step instructions for the assembly process. The HDT achieved success in all seven scenarios, providing accurate and clear guidance without the need for any user feedback. This phase highlighted the system's proficiency in delivering detailed and comprehensible instructions, ensuring users correctly executed each step. Here are some examples of the HDT's instructional responses utilizing our selected conversational paragraphs: **Inserting the Magazine:** User: "How exactly should the magazine be inserted and clicked into place?"; HDT: "Align the holes in the bottom of the barrel again and insert the short screw. Tighten it until it's secure but not overly tight." **Identifying Screws:** User: "I'm confused about which screw to use in the second step. How can I distinguish between the different screw sizes provided?"; HDT: "Locate the short screw—you should have only one of these, and it's the smallest screw provided."

**Phase 3: Recommendations/Solutions**
In the recommendations and solutions phase, the HDT offered corrective actions and alternative approaches to issues encountered during the assembly. The HDT achieved success in all 10 scenarios, effectively providing solutions without requiring feedback. This demonstrates the HDT's capability to proactively solve problems and assist users in overcoming assembly challenges. Here are some recommendations/solutions provided by the HDT for certain problems: **No Pre-drilled Holes:** User: "I'm currently in step 8 and it appears there are no pre-drilled holes for the screws."; HDT: "Align the Stock... Mark Screw Locations... Create Pilot Holes (Optional)... Insert the Screws..." **Skipping Steps:** User: "I believe the third step can be omitted."; HDT: "It's important to follow all the steps... Each step has a purpose... Skipping this step could misalign the barrel components..."

**Phase 4: Emotional Support**

The emotional support phase focused on detecting and responding to the user's visual and auditory emotional cues. The HDT achieved success in all 11 scenarios, providing appropriate encouragement and reassurance without the need for user feedback. This indicates the system's effectiveness in integrating metacognition input towards addressing the emotional and motivational needs of users, contributing to a positive user experience. During the emotional support phase, the HDT demonstrated effective responses to user emotions: **Reassurance and Encouragement:** User: "I'm feeling really frustrated with this assembly. I can't figure out step 7."; HDT: "It's completely natural... You're doing great, and you're not alone..."; **Step-by-Step Guidance:** User: "I don't understand step 6. Could you elaborate on it and then break it down into sub-steps?"; HDT: "Let's go back to Step 1... Insert Top Handguard... We'll tackle this together."; **Managing Emotions:** User: "This assembly is overwhelming. What if I fail in this step?"; HDT: "Take deep breaths... Step away for a few minutes..."; **Positive Reinforcement:** User: "That was a piece of cake. I'm proud of myself."; HDT: "That's awesome! You should feel proud, you're doing a fantastic job.... Great job pushing through!"

**Phase 5: Assembly Verification**

During the assembly verification phase, the HDT checked the final assembly for correctness. Out of six scenarios, the HDT achieved success in three, conditional success in another three, and no failures. Out of six scenarios, the HDT correctly verified the assembly in three scenarios on the first attempt: 5.2 (it seems like you are positioning the spring correctly on the bolt carrier), 5.4 (The dust cover in the second image appears to be somewhat misaligned), 5.6 (everything seems to be in place, including the magazine and display stand). For the remaining three scenarios, 5.1 (it's difficult to verify the exact placement of the screw in step 2), 5.3 (the muzzle appears to be installed correctly), 5.5 (Yes, the stock (J) is oriented correctly) the HDT needed one feedback instance to correct its response.

For the scenarios that required feedback, the HDT adjusted its responses based on additional input. For instance, when the user noted that the muzzle was loose, the HDT provided additional instructions to re-tighten and secure the muzzle properly. Similarly, when asked to check the alignment of the dust cover, the HDT corrected its initial response after the user pointed out a misalignment. This phase demonstrated the HDT's ability to verify assembly correctness and its capability to refine its responses based on user feedback, ensuring accurate and reliable guidance throughout the assembly process. Since it only needed one feedback attempt to give a correct response, it was quick to identify the mistake and correct it.

**Performance Dynamics of Interaction Modules**

The distribution of processing times for the interaction modules reveals significant insights into the architecture dynamics of the HDT system. The Speech Recognition Time, averaging $1.3 \pm 0.25$ seconds, exhibits minimal variability and constitutes a minor portion of the total interaction time. The Image Capture Time, with a mean of $2.13 \pm 0.32$ seconds, demonstrates moderate variation, contributing notably to the overall processing duration. The most time-consuming module is the LLM Response Time, averaging $9.72 \pm 2.9$ seconds, which represents a substantial proportion of the total interaction time. The 99th percentile of this distribution reaches 14.68 seconds, indicating significant fluctuations in processing duration. In contrast, the TTS Generation Time remains consistently low, with an average of $0.93 \pm 0.18$ seconds, contributing minimally to the total processing time. The total interaction pipeline commands an average processing time of $27.96 \pm 5.5$ seconds, with the 99th percentile of time consumption extending to 40.86 seconds. The high variability in the LLM Response Time and Image Capture Time significantly impacts the overall interaction duration. These findings highlight the need for optimization in the LLM processing and image capture modules to improve the efficiency and responsiveness of the HDT system.

**DISCUSSION**

Our results highlight strengths and areas for improvement in our HDT architecture when supporting HATs during a gun assembly task. The task performance of the HDT system as well as the architecture performance across different phases provides valuable insights into its capabilities and limitations. Our HDT prototype with the presented architecture demonstrated strong capabilities in real-time visual analysis and context-aware feedback during the pre-assembly and initial checks phase. Its ability to verify the presence of all required parts with minimal feedback indicates its potential to enhance task accuracy and efficiency. The system successfully identified missing parts after receiving feedback in scenarios where parts were deliberately removed, showcasing its adaptability and learning capability. This adaptability is crucial for ensuring all components are correctly identified and present before assembly begins, reducing the risk of errors.

During the instructional guidance phase, the HDT system excelled by providing accurate and clear step-by-step instructions for the assembly process. The 100% success rate in this phase underscores the system's proficiency in breaking down complex instructions into manageable sub-steps, ensuring users can follow along accurately without the need for feedback. This capability is vital for complex tasks requiring precise execution, as it helps users understand and correctly perform each step.

In the recommendations and solutions phase, the HDT's proactive problem-solving approach was evident. The system effectively offered corrective actions and alternative approaches to issues encountered during assembly, achieving a 100% success rate. This demonstrates the HDT's ability to assist users in overcoming assembly challenges, ensuring the task can proceed smoothly even when unexpected problems arise. The specific recommendations provided, such as addressing issues with pre-drilled holes or loose parts, highlight the system's practical problem-solving capabilities.

The HDT's effectiveness in providing emotional support was also notable. The system successfully detected and responded to users' emotional cues, offering appropriate encouragement and reassurance in all scenarios. This capability is essential for maintaining user motivation and confidence, particularly in tasks that can be stressful or frustrating. By addressing the emotional and motivational needs of users, the HDT contributes to a more positive and productive user experience.

The pre-assembly/initial checks and assembly verification phases revealed some limitations. While the HDT system achieved success in verifying parts and assembly correctness, it required feedback in several scenarios to correct its initial responses. The system's ability to adapt and correct mistakes based on user feedback indicates potential for continuous improvement. However, the need for more than one feedback attempt to verify parts highlights areas where the system's verification capabilities can be refined to achieve higher accuracy and reliability.

The analysis of interaction module processing times provides further insights into the system's performance dynamics. The speech recognition and text-to-speech generation times were relatively efficient, with minimal variability. However, the image capture and LLM response times significantly contributed to the overall interaction duration. The high variability in LLM response times suggests a need for optimization to enhance the system's responsiveness and efficiency. Reducing processing times in these modules would lead to more seamless and real-time interactions, further improving the user experience. Because our focus on a modular architecture, it is possible to work on the optimization or even replacement of each one of the modules without impacting the overall functionality of the HDT.

## CONCLUSION & FUTURE WORK

Our presented architecture and analysis demonstrate its feasibility and potential towards building HDTs for HATs through the integration of real-time visual analysis, context-aware feedback, problem-solving capabilities, emotional support, and physical embodiment. The system's success rates and adaptability shown in the performance analysis indicate its effectiveness in enhancing task accuracy and user experience. However, areas for improvement were identified in assembly verification accuracy and interaction module processing times, indicating the need for further optimization to enhance responsiveness and efficiency. Future implementations, including comprehensive user studies and custom domain-specific large language models, will be essential for enhancing the system's effectiveness, usability, and security, ultimately ensuring more seamless and efficient human-autonomy collaboration.

We identified three paths for future work. 1) to improve the overall system performance to accelerate all processing and computations to maintain uniform interactive rates independently of the complexity of the conversation and user emotional cues. We would like to experiment with different computational approaches, such as distributed computing as well as creating a custom LLM trained on domain-specific data. This tailored approach should allow the LLM to provide more precise and contextually relevant interactions, thereby improving the performance and reliability of the HDT system in specialized fields. By leveraging custom trained LLMs, we can better meet the unique needs of different applications, ensuring that our system remains adaptable and highly effective across various use cases; 2) to perform comprehensive user studies to evaluate our system in the context of HATs activities with diverse participants. These studies will provide valuable insights into the effectiveness, usability, and trust of the HDT system in various scenarios; 3) while our current architecture does not yet include data privacy and security measures, we recognize the importance of these aspects and plan to address them in future iterations. Areas to explore in this topic include avoid or provide customization on what user prompts and generated content are shared, even for situations where this is expected for model training purposes; incorporate content filtering and abuse monitoring to prevent harmful content generation. This ensures that data is processed securely and responsibly within the designated geography, maintaining compliance with data protection standards.

## ACKNOWLEDGMENTS

Special thanks to Dr. Roger Azevedo and Megan Wiedbusch for introducing us to metacognition and many fruitful discussions towards the next steps incorporating a metacognition architecture into our HDT.

**REFERENCES**

Ackerman, R., & Thompson, V. A. (2017). Meta-Reasoning: Monitoring and Control of Thinking and Reasoning. *Trends in Cognitive Sciences*, *21*(8), 607–617 .

Alawida, M., Mejri, S., Mehmood, A., Chikhaoui, B., & Isaac Abiodun, O. (2023). A Comprehensive Study of ChatGPT: Advancements, Limitations, and Ethical Considerations in Natural Language Processing and Cybersecurity. *Information*, *14*(8), Article 8.

Arya, R., Singh, J., & Kumar, A. (2021). A survey of multidisciplinary domains contributing to affective computing. *Computer Science Review*, *40*, 100399.

Dellermann, D., Calma, A., Lipusch, N., Weber, T., Weigel, S., & Ebel, P. (2021). *The future of human-AI collaboration: A taxonomy of design knowledge for hybrid intelligence systems* (arXiv:2105.03354). arXiv.

Gallagher, J., & Skalski, P. (2023, September 27). GPT-4 with Vision: Complete Guide and Evaluation. *Roboflow*. https://blog.roboflow.com/gpt-4-vision/

Guzman, K. G. (2024, February). Using GPT4 with Vision to tag and caption images. *OpenAI Cookbook*. https://cookbook.openai.com/examples/tag_caption_images_with_gpt4v

Hernandez, J., Suh, J., Amores, J., Rowan, K., Ramos, G. & Czerwinski, M.(2023, October) Affective Conversational Agents: Understanding Expectations and Personal Influences. https://www.microsoft.com/en-us/research/publication/affective-conversational-agents-understanding-expectations-and-personal-influences.

Hussain, S., Ameri Sianaki, O., & Ababneh, N. (2019). A Survey on Conversational Agents/Chatbots Classification and Design Techniques. In L. Barolli, M. Takizawa, F. Xhafa, & T. Enokido (Eds.), *Web, Artificial Intelligence and Network Applications* (pp. 946–956). Springer International Publishing.

Jiang, H., Zhang, X., Cao, X., Breazeal, C., Roy, D., & Kabbara, J. (2024). *PersonaLLM: Investigating the Ability of Large Language Models to Express Personality Traits* (arXiv:2305.02547). arXiv.

Jiang, N., Liu, X., Liu, H., Lim, E. T. K., Tan, C.-W., & Gu, J. (2022). Beyond AI-powered context-aware services: The role of human–AI collaboration. *Industrial Management & Data Systems*, *123*(11), 2771–2802.

Llanes-Jurado, J., Gómez-Zaragozá, L., Minissi, M. E., Alcañiz, M., & Marín-Morales, J. (2024). Developing conversational Virtual Humans for social emotion elicitation based on large language models. *Expert Systems with Applications*, *246*, 123261.

Manfredi, G., Erra, U., & Gilio, G. (2023). A Mixed Reality Approach for Innovative Pair Programming Education with a Conversational AI Virtual Avatar. *Proceedings of the 27th International Conference on Evaluation and Assessment in Software Engineering*, 450–454.

Microsoft. (n.d-b). *Windows.Media.SpeechRecognition* (Windows 11 Build 22621) Microsoft. Retrieved June 2024, from https://learn.microsoft.com/en-us/uwp/api/windows.media.speechrecognition?view=winrt-22621

Murdick, D. (2023, July). Building Trust in AI: A New Era of Human-Machine Teaming. *Center for Security and Emerging Technology*. https://cset.georgetown.edu/article/building-trust-in-ai-a-new-era-of-human-machine-teaming/

O'Neill, T., McNeese, N., Barron, A., & Schelble, B. (2022). Human–Autonomy Teaming: A Review and Analysis of the Empirical Literature. *Human Factors*, *64*(5), 904–938.

OpenAI. (2023). *GPT-4V(ision) system card*. https://cdn.openai.com/papers/GPTV_System_Card.pdf

OpenAI. (2024, May). *Hello GPT-4o*. Retrieved June 20, 2024, from https://openai.com/index/hello-gpt-4o/

Ray, P. P. (2023). ChatGPT: A comprehensive review on background, applications, key challenges, bias, ethics, limitations and future scope. *Internet of Things and Cyber-Physical Systems*, *3*, 121–154.

Roccas, S., Sagiv, L., Schwartz, S. H., & Knafo, A. (2002). The Big Five Personality Factors and Personal Values. *Personality and Social Psychology Bulletin*, *28*(6), 789–801.

Sandmann, S., Riepenhausen, S., Plagwitz, L., & Varghese, J. (2024). Systematic analysis of ChatGPT, Google search and Llama 2 for clinical decision support tasks. *Nature Communications*, *15*(1), 2050.

Shengli, W. (2021). Is Human Digital Twin possible? Computer Methods and Programs in Biomedicine Update, 1, 100014. https://doi.org/10.1016/j.cmpbup.2021.100014

So, C., Khvan, A., & Choi, W. (2023). Natural conversations with a virtual being: How user experience with a current conversational AI model compares to expectations. *Computer Animation and Virtual Worlds*, *34*(6), e2149.

Tankelevitch, L., Kewenig, V., Simkute, A., Scott, A. E., Sarkar, A., Sellen, A., & Rintel, S. (2024). The Metacognitive Demands and Opportunities of Generative AI. *Proceedings of the CHI Conference on Human Factors in Computing Systems*, 1–24.

*Text-to-Speech API* (Version 7). Unreal Speech. https://docs.unrealspeech.com/reference/getting-started-with-our-api

Wynne, K. T., & Lyons, J. B. (2018). An integrative model of autonomous agent teammate-likeness. *Theoretical Issues in Ergonomics Science*, *19*(3), 353–374.